\documentclass[a4paper,fleqn]{cas-dc}

\usepackage[numbers,sort&compress]{natbib}
\def\tsc#1{\csdef{#1}{\textsc{\lowercase{#1}}\xspace}}
\tsc{WGM}
\tsc{QE}
\begin{document}
\let\WriteBookmarks\relax
\def\floatpagepagefraction{1}
\def\textpagefraction{.001}

\shorttitle{}    

\shortauthors{Z. Fan et al.}  

\title [mode = title]{Nonreciprocal entanglement in cavity magnomechanics exploiting chiral cavity-magnon coupling}  


%\tnotetext[1]{Title footnote text} 


\author[1]{Zhiyuan Fan}%[<options>]

%\cormark[1]


%\ead{}


%\credit{Conceptualization of this study, Methodology, Software}

\affiliation[1]{organization={School of Physics, Zhejiang Key Laboratory of Micro-Nano Quantum Chips and Quantum Control, and State Key Laboratory for Extreme Photonics and Instrumentation},
            city={Hangzhou},
%          citysep={}, % Uncomment if no comma needed between city and postcode
            postcode={310027}, 
            state={Zhejiang},
            country={China}}

\author[1]{Xuan Zuo}%[]

% Footnote of the second author

% Email id of the second author
%\ead{}


%\credit{Methodology, Software}

\author[1]{Haotian Li}%[]
%\credit{Methodology, Software}

\author[1]{Jie Li}%[<options>]
\cormark[1]


\ead{jieli007@zju.edu.cn}


%\credit{Conceptualization of this study, Methodology, Supervision,}


\cortext[1]{Corresponding author}



\begin{abstract}
We propose a new mechanism to achieve nonreciprocal quantum entanglement in a cavity magnomechanical system by exploiting the chiral cavity-magnon coupling. The system consists of a magnon mode, a mechanical vibration mode, and two degenerate counter-propagating microwave cavity modes in a torus-shaped cavity. We show that nonreciprocal stationary microwave-magnon and -phonon bipartite entanglements and photon-magnon-phonon tripartite entanglement can be achieved by respectively driving different circulating cavity modes that hold a chiral coupling to the magnon mode. The nonreciprocal entanglements are shown to be robust against various experimental imperfections.  We specifically show how such nonreciprocal entanglement can realize the channel multiplexing quantum teleportation from a microwave field to a solid-state magnon mode. The work may find promising applications of the cavity magnomechanical systems in noise-tolerant quantum processing, channel multiplexing quantum teleportation, and chiral magnonic quantum networks.
\end{abstract}


%\item 
%\item 
%\item 

%\nocite{*}

% Keywords
\begin{keywords}
 Cavity magnonics\sep Cavity magnomechanics\sep Nonreciprocal entanglement\sep Chiral cavity-magnon coupling
\end{keywords}

\maketitle

% Main text
\section{Introduction}
 Hybrid systems based on collective spin excitations (magnons) in magnetic materials, e.g., yttrium iron garnet (YIG), have attracted considerable attention in the past decade. This is due to many excellent properties of the magnonic system, such as a large frequency tunability, a low dissipation rate, and an excellent ability to coherently couple with a variety of physical systems, including microwave photons \cite{Huebl2013,Tabuchi2014,Zhang2014}, optical photons \cite{Osada2016,Osada2018,ZhangX2016,Haigh2016}, vibration phonons \cite{Zhang2016,Li2018,Potts2021,Shen2022}, superconducting qubits \cite{Lachance-Quirion2017,Lachance-Quirion2020,Tabuchi2015,Xu2023}, {nitrogen-vacancy spins \cite{Hei2023,WangY2023},} etc.  As a subfield of hybrid magnonics, the couplings among microwave cavity photons, magnons, and magnetostriction-induced vibration phonons form the system of cavity magnomechanics (CMM) \cite{Zuo2023}. The CMM system provides a hybrid platform not only for studying strong interactions between light and matter, e.g., the triple magnon-photon-phonon strong coupling has been achieved~\cite{Shen23}; for studying rich nonlinear effects, such as Kerr-type nonlinearities~\cite{Shen2022}, magnonic frequency combs~\cite{Xiong23FR, Dong23}, synchronization~\cite{Cheng23}, non-Gaussianity and self-sustaining dynamics~\cite{Wen23}, but also for preparing many different kinds of quantum states, including entangled states \cite{Li2018,Li2019A,Li2020,Li2021B}, squeezed states \cite{Li2019B,ZhangW2021,Qian2023B,Li2023}, quantum steering \cite{Tan2019,ZhangW2022,Chen2021,Tan2023}, and quantum ground states of mechanical motion~\cite{Ding20,Lu21,Asjad23}.  These quantum states find potential applications in macroscopic quantum phenomena, quantum information processing, and quantum sensing \cite{Zuo2023,Lachance-Quirion2019,Yuan2022}.

Among many burgeoning research fields in advanced photonics, the chiral quantum optics \cite{Lodahl2017}, which exploits the new approaches and physical systems with chiral light-matter interaction \cite{Petersen2014,Kim2015,Dong2015,Gong2018}, finds many promising applications ranging from light technology to quantum information science \cite{Sollner2015,Mahmoodian2016}.  The chiral interaction can be achieved, e.g., by coupling the spin-momentum-locked light to quantum systems with polarization-dependent dipole transitions, such as atoms \cite{Mitsch2014,Pucher2022} and quantum dots \cite{Coles2016}. It is of significance to extend the advances in chiral optics to the microwave domain. In particular, along with the fast development of cavity magnonics \cite{Lachance-Quirion2019,Yuan2022,Bauer22}, the chiral microwave-magnon coupling has recently been proposed \cite{Yu2020} and demonstrated \cite{Bourhill2023} by employing a torus-shaped microwave cavity.  Besides, it has also been realized in the substrate-integrated-waveguide cavity \cite{Zhang2020}.

Here, we propose to achieve nonreciprocal quantum entanglement in CMM by exploiting the chiral cavity-magnon coupling.  We note that although nonreciprocal classical phenomena have been widely studied in various systems, nonreciprocal quantum effects have been much less explored. After the pioneering works of this kind~\cite{Huang2018,Jiao2020}, many nonreciprocal quantum protocols have been offered in cavity magnonics and the related field of optomechanics, including nonreciprocal entanglement \cite{Jiao2020,Yang2020,HJing22,Ren2022,Yang2023,Chakraborty2023,Chen2023B,Zheng2023,LYe,Liu2023,Zheng2024}, quantum steering \cite{Tan2019,ZhangW2022,Chen2021,Tan2023,Yang2021,Tan2021,Zhan2022,Guan2022,Zhong2023}, squeezing \cite{Xie2023,Guo2023}, photon \cite{Huang2018,LiB2019,Xu2020,XieH2022} and magnon \cite{You22} blockade, phonon \cite{Jiang2018,Xu2021} and magnon \cite{XuY2021,Huang2022} lasing, single-photon scattering \cite{Hafezi2012,Xu2017,Ren2022A,Xie2023A,Xiong23}, etc.   %Besides, nonreciprocal photon-pair quantum correlation~\cite{} and single-photon optical isolator~\cite{} have been experimentally demonstrated. %Besides quantum effects, nonreciprocal classical phenomena have been widely studied by exploiting different mechanisms, e.g., the optomechanical interaction \cite{Bernier2017,Peterson2017,Shen2016}, the interference between coherent and dissipative couplings \cite{Wang2019}, the chiral coupling \cite{Zhang2020,Bourhill2023}, the Kerr effect \cite{Cui2019}, the parity-time symmetry \cite{Peng2014,Peng2016}, the spinning induced frequency shift \cite{Maayani2018,Jiang2018,Xu2021}, etc.  
We note that most nonreciprocal entanglement proposals in cavity magnonics are based on a spinning microwave resonator configuration. However, unlike optical microcavities, it is still experimentally challenging to realize high-speed rotation of a large-size microwave cavity. Here, our proposal is based on the CMM system combining with the chiral cavity-magnon coupling realized in a torus-shaped microwave cavity, which has been demonstrated in the related experimental study \cite{Bourhill2023}. Specifically, we consider two degenerate counter-propagating microwave cavity modes, a magnon (Kittel) mode, and a mechanical vibration mode.  The chiral cavity-magnon coupling is achieved by placing the YIG sphere on a special line in the microwave cavity, where the Kittel mode couples solely to a specific circulating cavity mode and decouples from the other counter-propagating mode,  such that the time-reversal symmetry between the two degenerate modes is broken~\cite{Yu2020,Bourhill2023}. For the case of the cavity-magnon coupling being established, the effective magnomechanical coupling is greatly enhanced when the associated circulating cavity mode is strongly driven.  When the drive field is red-detuned from the magnon mode and the magnomechanical coupling becomes sufficiently strong, the magnon-phonon entanglement can be created, which further distributes to the circulating microwave mode, yielding the microwave-magnon and -phonon entanglements and the photon-magnon-phonon tripartite entanglement. For the case where the cavity-magnon coupling is absent, the magnomechanical subsystem is essentially decoupled from the (strongly driven) cavity. In this case, the magnomechanical dispersive coupling is too weak to generate any entanglement in the system.  Consequently, the chiral cavity-magnon coupling leads to the nonreciprocity of the microwave-magnon and -phonon entanglements and the tripartite entanglement when the two counter-propagating microwave cavity modes are respectively driven, e.g., via a waveguide.  We further analyze the effects of various imperfections in a real experiment, e.g., the backscattering induced coupling between the two degenerate modes and the inaccurate position of the YIG sphere in the cavity, which causes simultaneous couplings with the two degenerate modes, and find that the nonreciprocity of the entanglements is robust against these imperfections. We specifically show how a channel multiplexing quantum teleportation is possible based on the generated nonreciprocal entanglement, which manifests the potential application of our work in quantum information processing and chiral quantum networks.

The paper is organized as follows. In Sec.~\ref{syst}, we introduce the system, provide its Hamiltonian and the corresponding quantum Langevin equations (QLEs), and show how to obtain steady-state entanglements of the system. We then present the results of the nonreciprocal entanglements achieved in the ideal situation in Sec.~\ref{result}, and analyze the effects of some major experimental imperfections in Sec.~\ref{imperfect}. { We further discuss many potential applications of the proposed nonreciprocal entanglements in Sec.~\ref{appl}, and draw the conclusions in Sec.~\ref{conc}}.

\begin{figure}[h] 
	\centering
	\includegraphics[scale=0.24]{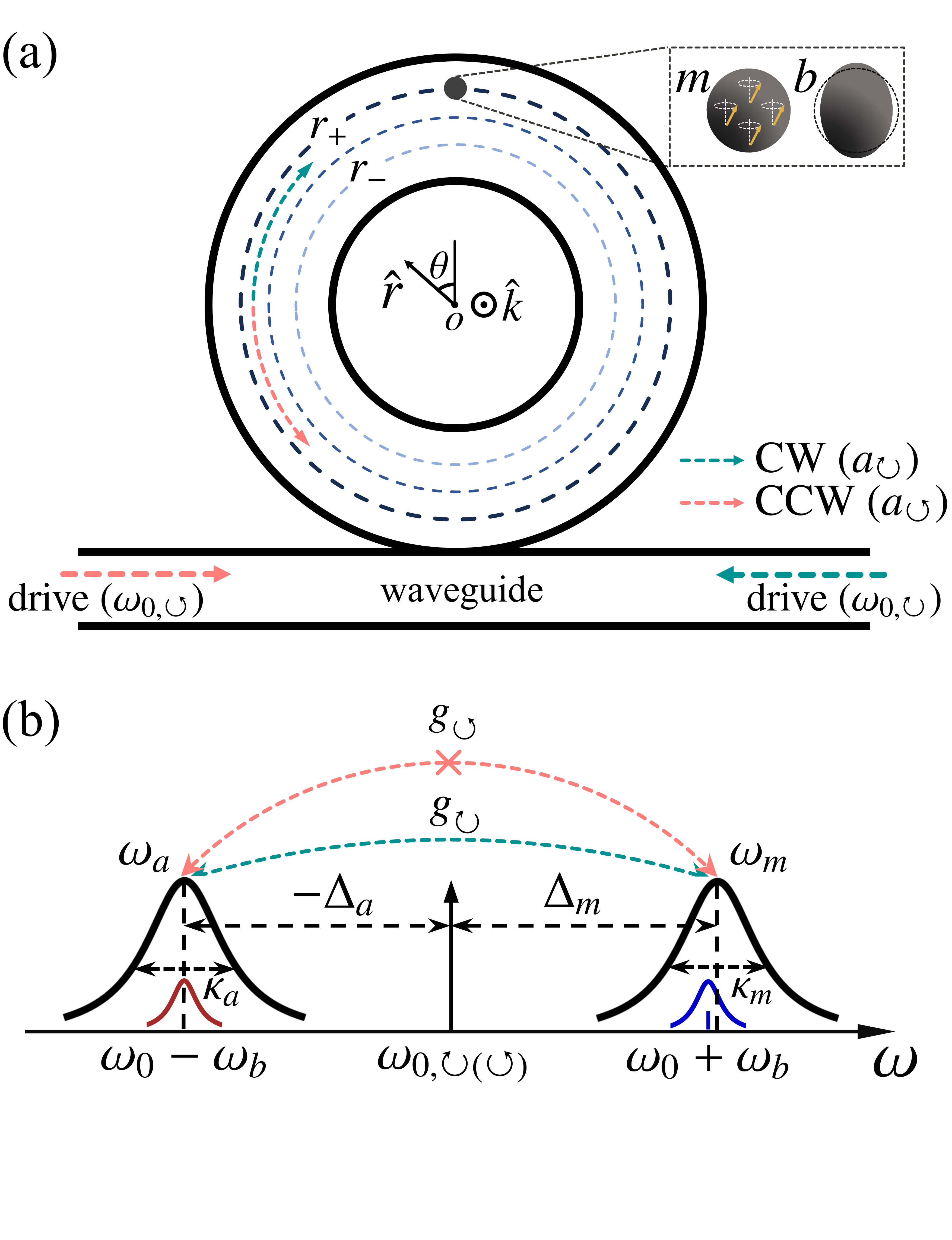}
	\caption{(a) Schematic diagram of the CMM system based on a torus-shaped microwave cavity. The CW (CCW) circulating microwave cavity mode is driven by an external microwave field at frequency $\omega_0$ via a waveguide.  $\hat{r}$ ($\hat{k}$) denotes the unit vector of the polar ($z$) axis. (b) Frequencies and linewidths of the system. Due to the chiral coupling between magnons and CW (CCW) propagating microwave photons, nonreciprocal microwave-magnon and -phonon bipartite entanglements and photon-magnon-phonon tripartite entanglement are generated when the cavity (magnon) mode is resonant with the Stokes (anti-Stokes) sideband scattered by the mechanical motion.}
	\label{fig1}
\end{figure}

\section{The Model}\label{syst}
The CMM system under study consists of a torus-shaped metallic microwave cavity \cite{Yu2020,Bourhill2023} and a ferrimagnetic sample, e.g., a YIG sphere, that is placed in the cavity; see Fig.~\ref{fig1}(a). The microwave cavity holds a pair of degenerate counter-propagating transverse electric (TE) microwave modes, which couple to a microwave waveguide through which a specific circulating mode is driven. The YIG sphere supports both a magnon (spin wave) mode and a mechanical vibration mode induced by the magnetostrictive interaction, which is a dispersive interaction for a large-size YIG sphere and provides necessary nonlinearity for creating entanglement in our protocol~\cite{Zuo2023}. The magnon mode further couples to the microwave cavity mode via the magnetic-dipole interaction. Specifically, we consider the magnon mode to be the Kittel mode, which corresponds to the uniform precession of all spins in the YIG sphere.
The Hamiltonian of the system reads
\begin{equation} \label{Hamiltonian}
	\begin{split}
		H/\hbar=&\sum_{j=\circlearrowright,\circlearrowleft}\omega_a a_j^\dagger a_j^{ }+\omega_m m^\dagger m+\frac{\omega_b}{2}(q^2+p^2)\\
		&+ \sum_{j=\circlearrowright,\circlearrowleft} g_j^{ } (a_j^\dagger m+a_j m^\dagger)+g_{m} m^\dagger m q+H_d/\hbar,
	\end{split}
\end{equation}
where $a_\circlearrowright^{ }$, $a_\circlearrowleft^{ }$ and $m$ ($a_\circlearrowright^\dagger$, $a_\circlearrowleft^\dagger$ and $m^\dagger$) denote the annihilation (creation) operators of the clockwise (CW) and counter-clockwise (CCW) circulating microwave cavity modes and the Kittel mode, respectively, and $q$ and $p$ are the dimensionless position and momentum of the mechanical motion, satisfying $[q,p]=i$. $\omega_a$, $\omega_m$, and $\omega_b$ are the resonance frequencies of the two degenerate cavity modes, the magnon mode, and the mechanical mode, respectively. %and the magnon frequency can be adjusted in a large range by varying the strength of the bias magnetic field. 
The cavity-magnon coupling $g_{\circlearrowright}^{ }$ ($g_\circlearrowleft^{ }$) describes the interaction strength between the magnons and the CW (CCW) circulating microwave photons. The single-magnon magnomechanical coupling rate $g_{m}$ is typically weak, but the effective magnomechanical coupling $G_m$ can be greatly enhanced by directly driving the magnon mode \cite{Shen23}, or driving the microwave cavity via the cavity-magnon excitation-exchange interaction \cite{Zhang2016,Potts2021,Shen2022}. The last term $H_d/\hbar=i E(a_l^\dagger e^{-i\omega_0 t}-a_l e^{i\omega_0 t})$ describes the microwave driving of the CW ($l=\circlearrowright$) or CCW ($l=\circlearrowleft$) circulating cavity mode, where $E=\sqrt{2\kappa_{a,e} P_0/\hbar\omega_0}$ is the coupling strength between the cavity mode and the drive field, with $\kappa_{a,e}$ being the external dissipation rate of the cavity and $P_0$ ($\omega_0$) the power (frequency) of the drive field.

The symmetry between the CW and CCW circulating microwave modes in coupling with the Kittel mode can be manipulated to be broken by placing the YIG sphere on a special line in the torus-shaped cavity, which leads to the chiral cavity-magnon coupling. %i.e., $g_\circlearrowright^{ }\neq g_\circlearrowleft^{ }$. 
This is due to the fact that the Kittel mode magnetization precesses anticlockwise around the effective magnetic field and couples to the microwave photons with the same polarization, and that the polarization of some cavity modes depends on their propagation direction at special lines in the cavity~\cite{Yu2020}.   To be specific, the Kittel mode only couples to the right-handed circularly polarized magnetic field with respect to the $z$ axis. For a TE mode, this corresponds to a CW (CCW) mode for the YIG sphere being positioned on the special line of, e.g., $r=r_+$ ($r=r_-$), cf. Fig.~\ref{fig1}(a), where $r$ is the radial coordinate \cite{Bourhill2023}.  Accordingly, the Kittel mode decouples to the counter-propagating cavity mode on the same circular line.
%At special circulating line in the microwave cavity, e.g., the YIG sphere position $\{\rho,\phi\}=\{\rho_+,0\}$ as depicted in Fig. \ref{fig1}(a), the magnetic components $H_\phi$ and $H_\rho$ of the counter-propagating CW and CCW circulating modes respectively oscillate $\pm 90^\circ$ out of phase with each other ($H_\phi=iH_\rho$ for CW mode and $H_\phi=-iH_\rho$ for CCW mode), which correspond to the orthogonal circular polarizations and opposite chiralities \cite{Yu2020,Bourhill2023}. Analogous to the chiral coupling of quantum systems with polarization-dependent dipole transitions and spin-momentum locked light \cite{Lodahl2017}, the Kittel magnon mode in the ferrimagnet YIG sphere describes the uniform spin precession around the effective magnetic field in a unified manner (with magnetization component $M_\phi=iM_\rho$), and on the physical perspective it tends to couple to the CW circulating cavity mode with the same chirality, which leads to the effective cavity-magnon coupling $g_\circlearrowright^{ }\neq 0$. While on the other hand, the CCW circulating microwave mode behaves completely decoupled with the magnon mode ($g_\circlearrowleft^{ }=0$) \cite{Bourhill2023}, cf. Fig. \ref{fig1}(b). 
As a result, the chiral cavity-magnon coupling is achieved. The chiral coupling shows good tunability: %e.g., the position $\{\rho,\phi\}=\{\rho_-,0\}$, the polarizations and chiralities of CW and CCW modes get reversed with respect to the case as discussed at $\{\rho,\phi\}=\{\rho_+,0\}$, such that 
one can realize the cavity-magnon couplings taking either $g_\circlearrowright^{ }>0$ and $g_\circlearrowleft^{ }= 0$, or $g_\circlearrowleft^{ }> 0$ and $g_\circlearrowright^{ }=0$, by placing the YIG sphere on a specific orbit.

Specifically, in our protocol we consider the chiral coupling for the YIG sphere being positioned at, e.g., $r=r_+$, and we thus have $g_\circlearrowright^{ }>0$ and $g_\circlearrowleft^{ }=0$. By including the dissipation and input noise of each mode and working in the interaction picture with respect to $\hbar \omega_0(a_\circlearrowright^\dagger a_\circlearrowright^{ }+a_\circlearrowleft^\dagger a_\circlearrowleft^{ }+m^\dagger m)$, we obtain the following QLEs of the system, in the case of the CW mode being strongly driven:
\begin{equation} \label{QLEs}
	\begin{split}
		\dot{a}_\circlearrowright^{ }=&-i\Delta_{\circlearrowright} a_\circlearrowright^{ }-\kappa_a a_\circlearrowright^{ }-i g_\circlearrowright^{ } m+ E+\sqrt{2\kappa_a}a_{\circlearrowright}^{\mathrm{in}},\\
		\dot{a}_\circlearrowleft^{ }=&-i\Delta_{\circlearrowleft} a_\circlearrowleft^{ }-\kappa_a a_\circlearrowleft^{ }+\sqrt{2\kappa_a}a_{\circlearrowleft}^{\mathrm{in}},\\
		\dot{m}=&-i\Delta_m m-\kappa_m m-i g_\circlearrowright^{ } a_\circlearrowright^{ }-i g_m mq+\sqrt{2\kappa_m}m^{\mathrm{in}},\\
		\dot{q}=&\ \omega_b p,\ \ 
		\dot{p}=-\omega_b q-\gamma_b p-g_m m^\dagger m+\xi,\\
	\end{split}
\end{equation}
where $\Delta_{\circlearrowright (\circlearrowleft)} \equiv \Delta_{a}=\omega_a-\omega_0$, $\Delta_m=\omega_m-\omega_0$,  $\kappa_m$ ($\gamma_b$) is the dissipation rate of the magnon (mechanical) mode, and $\kappa_a=\kappa_{a,i}+\kappa_{a,e}$ represents the total dissipation rate of the microwave cavity, with $\kappa_{a,i}$ being the internal dissipation rate. Here, $k^{\mathrm{in}}$ ($k=a_\circlearrowright^{ }$, $a_\circlearrowleft^{ }$, $m$) denote the zero-mean input noises of the CW, CCW, and magnon modes, respectively. They obey the following correlation functions: $\langle k^{\mathrm{in}}(t) k^{\mathrm{in}\dagger}(t') \rangle=[N_{k}(\omega_k)+1]\delta(t-t')$ and $\langle k^{\mathrm{in}\dagger}(t) k^{\mathrm{in}}(t') \rangle=N_{k}(\omega_k)\delta(t-t')$.  The Brownian noise operator $\xi(t)$ is for the mechanical oscillator, which is intrinsically non-Markovian. Nevertheless, for a high mechanical quality factor $Q_b=\omega_b/\gamma_b \gg 1$, the Markovian approximation can be made \cite{Giova2001}, leading $\xi(t)$ to be $\delta$-correlated: $\langle \xi(t)\xi(t')+\xi(t')\xi(t)\rangle/2\simeq \gamma_b[2N_b(\omega_b)+1] \delta(t-t')$. In the above noise correlations, $N_j(\omega_j)=[\mathrm{exp}(\hbar\omega_j/k_B T)-1]^{-1}$ ($j=a$, $m$, $b$) is the mean thermal excitation number of the corresponding mode at bath temperature $T$, with $k_B$ being the Boltzmann constant.

The generation of entanglement in the present system requires a sufficiently strong magnomechanical coupling strength. To this end, we strongly drive the CW cavity mode via a microwave waveguide (Fig.~\ref{fig1}(a)). Due to the cavity-magnon excitation-exchange interaction, the magnon mode also gets effectively pumped, leading to $|\langle m \rangle| \gg 1$. This allows us to linearize the system dynamics around the large average values by writing each operator as $O=\langle O \rangle +\delta O$ ($O\, {=}\, a_\circlearrowright^{ }, a_\circlearrowleft^{ }, m,q,p$) and neglecting small second-order fluctuation terms. Consequently, we obtain a set of linearized QLEs for the quantum fluctuations of the system $(\delta a_\circlearrowright^{ }, \delta a_\circlearrowleft^{ }, \delta m, \delta q, \delta p)$, which can be rewritten in the quadrature form as follows
\begin{equation} \label{matrixform}
	\dot{u}(t)=Au(t)+n(t),
\end{equation}
where $u(t)=[\delta X_{a_\circlearrowright}(t),\delta Y_{a_\circlearrowright}(t),\delta X_{a_\circlearrowleft}(t),\delta Y_{a_\circlearrowleft}(t),\delta X_m(t),$ $\delta Y_m(t),\delta q(t),\delta p(t)]^{T}$ denotes the vector of the quadrature fluctuations of the system, and $\delta X_{k}=(\delta k+\delta k^\dagger)/\sqrt{2}$ and $\delta Y_k=i(\delta k^\dagger-\delta k)/\sqrt{2}$ ($k=a_\circlearrowright^{ }$, $a_\circlearrowleft^{ }$, $m$), and $n(t)=[\sqrt{2\kappa_a}X_{a_\circlearrowright}^{\mathrm{in}}(t), \sqrt{2\kappa_a}Y_{a_\circlearrowright}^{\mathrm{in}}(t), \sqrt{2\kappa_a}X_{a_\circlearrowleft}^{\mathrm{in}}(t), \sqrt{2\kappa_a}Y_{a_\circlearrowleft}^{\mathrm{in}}(t),\sqrt{2\kappa_m}$ $X_m^{\mathrm{in}}(t), \sqrt{2\kappa_m}Y_m^{\mathrm{in}}(t),0,\xi(t)]^{T}$ is the vector of the input noises. The drift matrix $A$ is given by
\begin{small}
	\begin{equation} \label{drift}
		\begin{split}
			&A{=}\\
			&\begin{pmatrix}
				-\kappa_a & \Delta_\circlearrowright & 0 & 0 & 0 & g_\circlearrowright^{ } & 0 & 0\\
				-\Delta_\circlearrowright & -\kappa_a & 0 & 0 & -g_\circlearrowright^{ } & 0 & 0 & 0\\
				0 & 0 & -\kappa_a & \Delta_\circlearrowleft & 0 & 0 & 0 & 0\\
				0 & 0 & -\Delta_\circlearrowleft & -\kappa_a & 0 & 0 & 0 & 0\\
				0 & g_\circlearrowright^{ } & 0 & 0 & -\kappa_m & \tilde{\Delta}_m & \mathrm{Im}G_m  & 0\\
				-g_\circlearrowright^{ } & 0 & 0 & 0 & -\tilde{\Delta}_m & -\kappa_m & -\mathrm{Re}G_m & 0\\
				0 & 0 & 0 & 0 & 0 & 0 & 0 & \omega_b\\
				0 & 0 & 0 & 0 & -\mathrm{Re}G_m & -\mathrm{Im}G_m & -\omega_b & -\gamma_b
			\end{pmatrix},
		\end{split}
	\end{equation}
\end{small}where the effective magnomechanical coupling strength $G_m=\sqrt{2}g_{m}\langle m\rangle$, which is greatly enhanced due to the large amplitude of the magnon mode under a strong drive field, i.e., 

\begin{equation}\label{mave}
	\langle m\rangle=\frac{-ig_\circlearrowright^{ }}{\kappa_m+i\tilde{\Delta}_m}\langle a_{\circlearrowright}\rangle =\frac{-ig_\circlearrowright^{ } E}{g_{\circlearrowright}^2+(\kappa_a+i\Delta_a)(\kappa_m+i\tilde{\Delta}_m)},
\end{equation} 

The effective magnon-drive detuning $\tilde{\Delta}_m=\Delta_m+g_m\langle q\rangle$ includes the frequency shift due to the magnomechanical dispersive interaction, where the steady-state mechanical displacement $\langle q\rangle=-g_m|\langle m\rangle|^2/\omega_b$. 
%there will be completely no effective activating of CW circulating microwave photons and magnons (i.e., $\langle a_\circlearrowright^{ }\rangle=\langle m\rangle=0$) due to the decoupling between the pumped CCW circulating cavity mode and the magnon mode, which ultimately gives rise to the suppression of the effective magnemechanical coupling strength. As a consequence, in this case, the drift matrix $A$ takes the same form as in Eq.~\eqref{drift}, but with $G_m=0$, and such that, the key magnomechanical nonlinear interaction in the hybrid cavity magnomechanical system to generate entanglement is of absence in this case.

Because of the linearized dynamics and the Gaussian nature of the input noises, the steady state of the quantum fluctuations of the system retains a Gaussian state, which can be described by an $8\times8$ covariance matrix (CM) $V$, with its entries defined as $V_{ij}=\langle u_i(t)u_j(t')+u_j(t')u_i(t) \rangle/2$ ($i,j\in\{1,2,...,8\}$). At the steady state, the CM can be conveniently achieved by solving the Lyapunov equation~\cite{Vitali2007}
\begin{equation}
	AV+VA^T=-D,
\end{equation}
where $D=\mathrm{diag}[\kappa_a(2N_a+1),\kappa_a(2N_a+1),\kappa_a(2N_a+1),\kappa_a(2N_a+1),$ $\kappa_m(2N_m+1),\kappa_m(2N_m+1),0,\gamma_b(2N_b+1)]$ is the diffusion matrix, which is defined via $D_{ij}\delta(t-t')=\langle n_i(t)n_j(t')+n_j(t')n_i(t) \rangle/2$. When the CM is obtained, we adopt the logarithmic negativity $E_N$ to calculate the degree of any bipartite entanglement of the system, which is defined as \cite{Adesso2004}
\begin{equation}
	E_N=\mathrm{max}[0,-\mathrm{ln}(2\eta^-)],
\end{equation}
where $\eta^-\equiv 2^{-1/2}[\Sigma-(\Sigma^2-4\mathrm{det} V_4)^{1/2}]^{1/2}$ and $V_4$ is the $4\times 4$ CM of the subsystem of the modes $e$ and $f$ ($e,f=a_\circlearrowright^{ }$, $a_\circlearrowleft^{ }$, $m$, $b$), with $V_e$, $V_f$, $V_{ef}$ and $V_{fe}$ being $2\times2$ blocks of $V_4$, and $\Sigma=\mathrm{det} V_e+\mathrm{det} V_f-2\mathrm{det} V_{ef}$. A nonzero $E_N$ indicates that the associated bipartite system is entangled.

\section{Nonreciprocal entanglements in the ideal case}\label{result}
Here we present the results of the nonreciprocal entanglements achieved in the system. The nonreciprocal entanglements refer to the fact that, by strongly driving the CW circulating cavity mode, entanglements, e.g., microwave-magnon and -phonon entanglements, can be generated, while they are absent when driving the counter-propagating CCW mode, cf. Fig.~\ref{fig1}(a).  This corresponds to the ideal situation without considering any experimental imperfections. In the next section, we shall study the effects of various imperfections on the achieved nonreciprocal entanglements.

\begin{figure}[h] 
	\centering
	\includegraphics[width=0.95\linewidth]{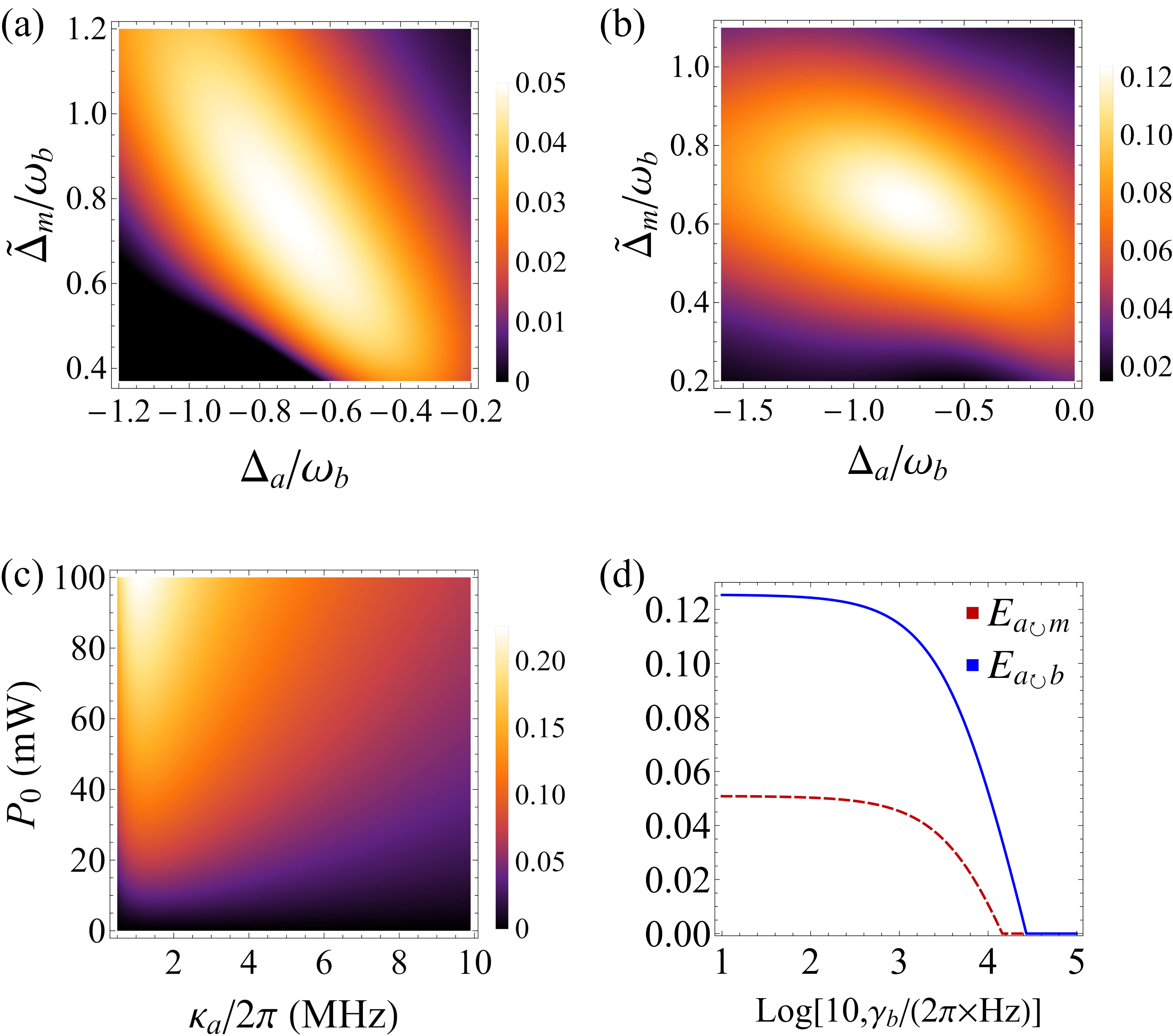}
	\caption{Density plot of the entanglement (a) $E_{a_{\circlearrowright}m}$ and (b) $E_{a_{\circlearrowright}b}$ versus detunings $\Delta_a$ and $\tilde{\Delta}_m$. (c) Density plot of $E_{a_{\circlearrowright}b}$ versus cavity decay rate $\kappa_a$ and drive power $P_0$ at optimal detunings $\Delta_a=-0.76\omega_b$ and $\tilde{\Delta}_m=0.65\omega_b$.  $\kappa_{a,i}$ is fixed in varying $\kappa_a$. The other parameters are the same as in (b). (d) Microwave-magnon entanglement $E_{a_{\circlearrowright}m}$ (dashed) and microwave-phonon entanglement  $E_{a_{\circlearrowright}b}$ (solid) versus mechanical damping rate $\gamma_b$. The parameters used for plotting $E_{a_{\circlearrowright}m}$ ($E_{a_{\circlearrowright}b}$) are as the same in (a) [(b)] but at optimal detunings $\Delta_a=-0.72\omega_b$ and $\tilde{\Delta}_m=0.76\omega_b$ ($\Delta_a=-0.76\omega_b$ and $\tilde{\Delta}_m=0.65\omega_b$). %Here we adopt the $E_{\mathrm{am}}$ and $E_{\mathrm{ab}}$ to replace $E_{a_{\circlearrowright}m}$ and $E_{a_{\circlearrowright}b}$ in (d) for simplicity. 
		See text for the other parameters. } 
	\label{fig2}
\end{figure}

We start with the case where the CW mode is driven. The CCW mode is essentially decoupled from the tripartite system of the CW, magnon, and mechanical modes, because of $g_\circlearrowright^{ }>0$ and $g_\circlearrowleft^{ }=0$. The system is then reduced to the tripartite CMM system and
%In order to achieve the nonreciprocal entanglement based on the distinct chiral cavity-magnon coupling about the two counter-propagating microwave modes in this torus-shaped cavity magnomechanical configuration, e.g., the entanglement can only be generated by effectively driving the CW circulating cavity mode, but not the CCW circulating one, here we mainly consider that the ferrimagnet YIG sphere is placed at $\{\rho,\phi\}=\{\rho_+,0\}$ as shown in Fig. \ref{fig1}(a), which provides the significant asymmetric cavity-magnon coupling that $g_{\circlearrowright}^{ }\gg g_{\circlearrowleft}^{ }=0$. The underlying physics of entanglement generation in this scheme, corresponding to the entanglement mechanism as used in the general cavity magnomechanical system 
we adopt a red-detuned microwave drive field with respect to the magnon mode with detuning $\tilde{\Delta}_m \approx \omega_b$ (Fig.~\ref{fig1}(b)), which is responsible for significantly cooling the low-frequency mechanical mode to eliminate the dominant thermal noise entering the system. The cooling interaction optimally works in the resolved-sideband limit of $\kappa_m \ll \omega_b$, which is well satisfied for the CMM experiments using YIG~\cite{Zhang2016,Potts2021,Shen2022,Shen23}. A precondition for seeing entanglement in the system is that the mechanical mode should be cooled into its quantum ground state~\cite{Li2018,Li2019A,Li2020}. However,  with only the effective magnomechanical cooling (beam-splitter) interaction, it is insufficient to produce entanglement. The magnomechanical parametric down conversion (PDC) interaction should be activated. Typically, this is realized by driving the magnon mode with a blue-detuned microwave field, which, however, can easily cause the system to be unstable. We thus adopt a {\it sufficiently strong} red-detuned drive field~\cite{Vitali2007}, such that the weak-coupling condition $G_m \ll \omega_b$ for taking the rotating-wave (RW) approximation to obtain the magnomechanical beam-splitter interaction $\propto m^\dagger b+ m b^\dagger$ is broken, where $b$ denotes the mechanical mode and $b=(q+ip)/\!\sqrt{2}$.  Consequently, the counter-RW terms, which correspond to the PDC interaction $\propto m^\dagger b^\dagger+ m b$, start to play the role in creating magnomechanical entanglement. The entanglement is further distributed to the circulating cavity mode when the cavity is resonant with the Stokes mechanical sideband, i.e., $\Delta_a \approx -\omega_b$, yielding the entanglement between the CW microwave mode with the magnon (mechanical) mode, as shown in Fig.~\ref{fig2}(a) (Fig.~\ref{fig2}(b)).  The entanglements are maximized around the optimal detunings as depicted in Fig.~\ref{fig1}(b), i.e., the CW cavity and magnon modes are respectively resonant with the Stokes and anti-Stokes sidebands of the drive field.

%	In contrast, for the case that the CCW circulating mode experiences completely same microwave driving as used in entanglement protocol we proposed above, because of the decoupling between the CCW circulating cavity mode and magnon mode, clearly, the magnon mode will not obtain effective activation, and this leads to the absence of effective magnomechanical nonlinear interaction for entanglement generation. As a result, there will be no any bipartite entanglement in this case, 

%	In view of the physical mechanism of the quantum entanglement, the foremost task of studying such a hybrid system is to show the nonreciprocal bipartite entanglement. Here we display the stationary CW circulating microwave photon-magnon entanglement $E_{a_{\circlearrowright}m}$ and -phonon entanglement $E_{a_{\circlearrowright}b}$ versus the detunings $\Delta_a$ and $\tilde{\Delta}_m$ in Figs. \ref{fig2}(a) and \ref{fig2}(b) by assuming that the CW circulating cavity mode gets effective pumping. 

%However, when we reconsider driving the CCW circulating cavity mode in this hybrid system, the quantum entanglement is clearly absent without effective activating the  magnons and the mechanical vibration phonons, which provide the indispensable nonlinear magnomechanical interaction to yield the quantum effect. 

We have employed feasible parameters \cite{Zhang2016,Li2018,Potts2021,Shen2022,Shen23,Bourhill2023}: $\omega_a/2\pi\approx\omega_m/2\pi\approx  \omega_0/2\pi =10$ GHz, $\omega_b/2\pi=10$ MHz, $\kappa_m/2\pi=1$ MHz, $\gamma_b/2\pi=10^2$ Hz, $\kappa_{a,i}/2\pi=0.2$ MHz, $\kappa_{a,e}/2\pi=2.8$ MHz ($4.8$ MHz), $g_\circlearrowright^{ }/2\pi=4$ MHz ($8$~MHz), and $T=10$ mK in getting Fig.~\ref{fig2}(a) (Fig.~\ref{fig2}(b)). 
These parameters lead to a complex steady-state average of $\left\langle m\right\rangle$, cf. Eq.~\eqref{mave}. Therefore, Fig.~\ref{fig2}(a) (Fig.~\ref{fig2}(b)) is plotted with a fixed
amplitude of the effective coupling $|G_m|/2\pi=4.0$ MHz ($2.5$~MHz). Figure~\ref{fig2}(c) shows the drive power required to generate considerable microwave-phonon entanglement for different values of the cavity decay rate. Clearly, one can increase the drive power to compensate the cavity losses in such a torus-shaped microwave cavity.   Figure~\ref{fig2}(d) plots the entanglements $E_{a_{\circlearrowright}m}$ and $E_{a_{\circlearrowright}b}$ versus the mechanical damping rate $\gamma_b$. Both the entanglements can survive even for a very large damping rate $\gamma_b/2\pi\sim10^4$ Hz, corresponding to the mechanical quality factor $Q_b \sim 10^3$, which can be easily achieved in the CMM experiments~\cite{Zuo2023}. Note that the results shown in Fig.~\ref{fig2} are in the steady state, which is guaranteed by the negative eigenvalues (real parts) of the drift matrix $A$. Under the parameters of Fig. \ref{fig2}(c) and taking $\kappa_a$, e.g. as used in Fig. \ref{fig2}(b), the maximum coupling strength $|G_m|$ for keeping the system stable is $2\pi\times 11.9$ MHz, which is much stronger than the value we used, so the system remains stable. Besides, a strong drive might lead to magnon frequency combs \cite{Xiong23FR}, which we check below. With the parameters of Fig. \ref{fig2}(c) and $\kappa_a$  used in Fig. \ref{fig2}(b), by  investigating the time evolution of each mode of the system (Appendix A), we determine that the critical threshold of the coupling strength $|G_m|$ for producing the magnon frequency combs is about $2\pi\times 8.5$ MHz, which corresponds to a microwave driving power of $0.9$ W. The drive power we used is far below this value, so the presence of magnon frequency combs can be excluded. Note that the threshold power looks high, because we take a feasible relatively large cavity decay rate.

\begin{figure}[h] 
	\centering
	\includegraphics[width=1\linewidth]{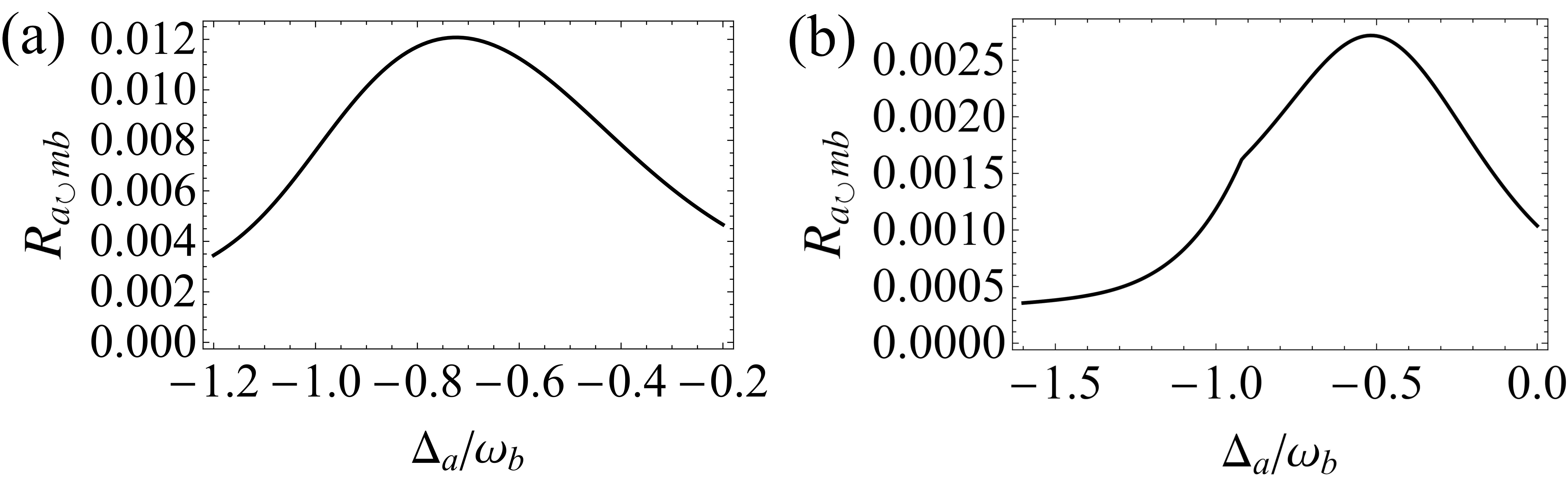}
	\caption{(a) [(b)]Tripartite entanglement in terms of the minimum residual contangle  $R_{a_{\circlearrowright}mb}$ versus $\Delta_a$, with the paramters as used in plotting of Fig. \ref{fig2}(a) [Fig. \ref{fig2}(b)] and $\tilde{\Delta}_m=0.76\omega_b$ ($0.65\omega_b$).}
	\label{fig3}
\end{figure}

Apart from bipartite entanglements studied above, we confirm that the genuine photon-magnon-phonon tripartite entanglement is also present in our system. The genuine tripartite entanglement is quantified by the minimum residual contangle $R_\tau^{\mathrm{min}}$~\cite{Adesso2007}, which can be conveniently computed when the CM is achieved \cite{Li2018}. Figs. \ref{fig3}(a) and \ref{fig3}(b) respectively show the tripartite entanglement measure  $R_{a_{\circlearrowright}mb}$ versus the detuning $\Delta_a$, under the parameters as used in Figs. \ref{fig2}(a) and \ref{fig2}(b).

For the case where the CCW cavity mode is driven, due to the absence of the CCW mode-magnon coupling, i.e., $g_\circlearrowleft^{ }=0$, the magnon mode is effectively not driven, and hence the dispersive magnomechanical coupling is too weak to create any entanglement~\cite{Zuo2023}, i.e., $E_{a_{\circlearrowleft}m}=E_{a_{\circlearrowleft}b}=0$ and $R_{a_{\circlearrowleft}mb}=0$.  Therefore, significant nonreciprocal microwave-magnon, microwave-phonon and microwave-magnon-phonon entanglements are achieved by exploiting the chiral cavity-magnon coupling in the system.

\begin{figure}[h] 
	\centering
	\includegraphics[width=1\linewidth]{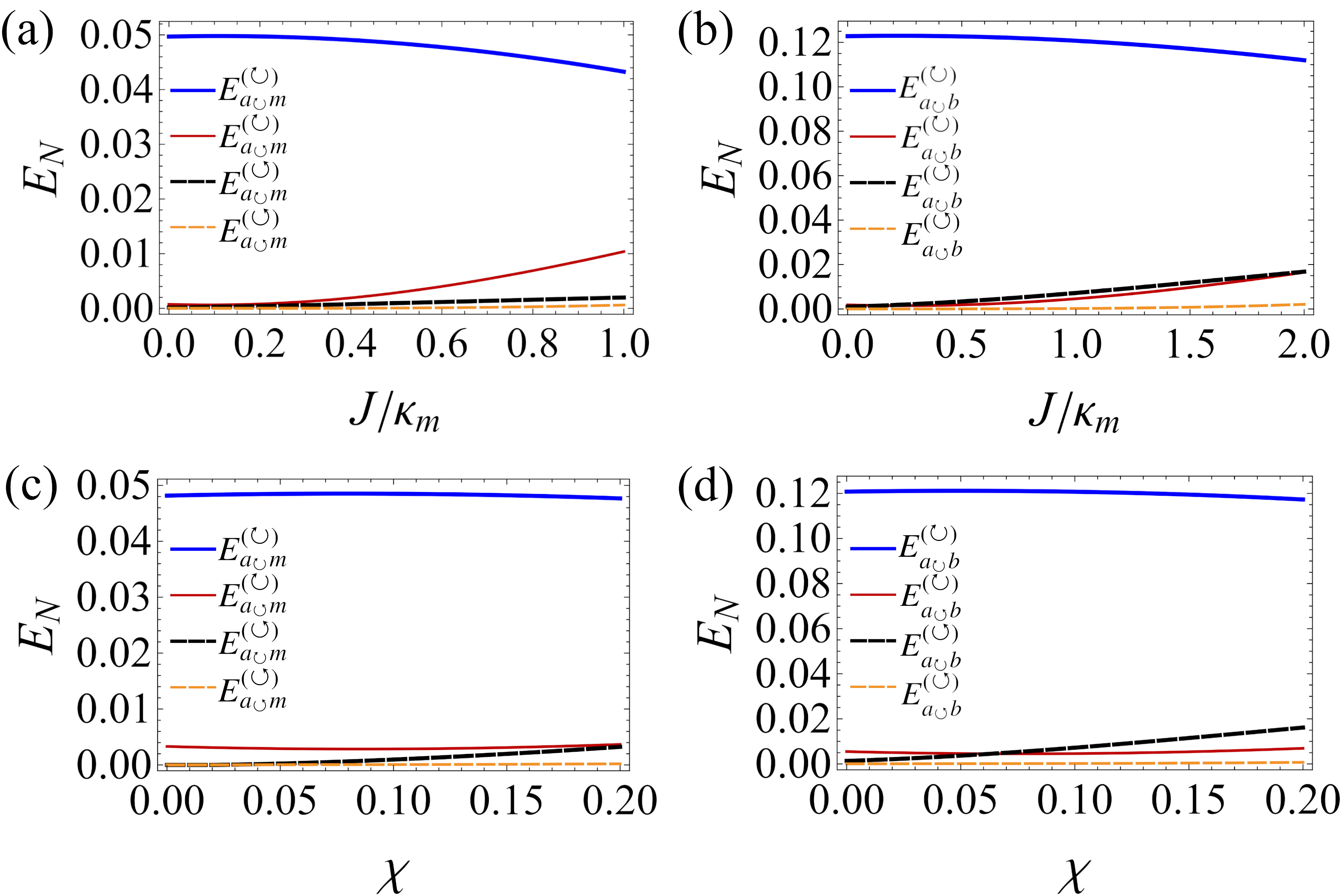}
	\caption{Stationary entanglement $E_{a_{\circlearrowright}\nu}^{(\circlearrowright)}$ (thick solid), $E_{a_{\circlearrowleft}\nu}^{(\circlearrowright)}$ (thin solid), $E_{a_{\circlearrowright}\nu}^{(\circlearrowleft)}$ (thick dashed) and $E_{a_{\circlearrowleft}\nu}^{(\circlearrowleft)}$ (thin dashed) versus coupling strength $J$ under $\chi \equiv g_\circlearrowleft^{ }/g_\circlearrowright^{ }=0.1$ in (a)-(b); versus $\chi$ under $J=0.5\kappa_m$ in (c) and $J=\kappa_m$ in (d). Here, the subscript $\nu=m$ in (a) and (c) for microwave-magnon entanglement; $\nu=b$ in (b) and (d) for microwave-phonon entanglement. The superscript ($\circlearrowright$) or ($\circlearrowleft$) denotes the specific circulating mode under drive. The drive power used in (a) and (c) [(b) and (d)] corresponds to the value of $|G_m|$ used for $E_{a_{\circlearrowright}m}$ ($E_{a_{\circlearrowright}b}$) in Fig.~\ref{fig2}(d). We take $\gamma_b/2\pi=10^2$ Hz, and the other parameters in (a) and (c) [(b) and (d)] are the same as those used to obtain $E_{a_{\circlearrowright}m}$ ($E_{a_{\circlearrowright}b}$) in Fig.~\ref{fig2}(d). 		}
	\label{fig4}
\end{figure}

\section{Nonreciprocal entanglements versus experimental imperfections}\label{imperfect}

In an actual experiment, various imperfections may cause appreciable effects on the nonreciprocity of the entanglements achieved in the ideal case (Sec.~\ref{result}). These include the coupling between the two degenerate counter-propagating cavity modes by backscattering from the YIG sphere or the cavity ports \cite{Bourhill2023}, which can be modeled as $\hbar J(a_\circlearrowright^{ } a_\circlearrowleft^\dagger + a_\circlearrowright^\dagger a_\circlearrowleft^{ })$ with $J$ being the coupling strength, and the residual coupling of the Kittel mode to the CCW circulating mode, i.e., $ g_\circlearrowleft^{ } \ll g_\circlearrowright^{ }$ but $g_\circlearrowleft^{ } \neq 0$, due to, e.g., the inaccurate position of the YIG sphere in the cavity. The inclusion of these two effects leads to a new Hamiltonian of the system and linearized QLEs for the quantum fluctuations, which are provided in the Appendix B for brevity.

Figure~\ref{fig4} manifests the impacts of the above two imperfections on the nonreciprocity of the entanglements. Specifically, in Fig.~\ref{fig4}(a) we show the nonreciprocal microwave-magnon entanglement as a function of the mode coupling strength $J$ under $\chi \equiv g_\circlearrowleft^{ }/g_\circlearrowright^{ }=0.1$ ($g_\circlearrowright^{ }$ is fixed as in Fig.~\ref{fig2}(a)). %Note that in this case, the drift matrix takes the renewed form which includes $J$ and $g_\circlearrowleft^{ }$, as can be seen in the Appendix. 
The four curves correspond, respectively, to the entanglement between the magnon mode with the CW or the CCW mode when either the CW or the CCW mode is driven. For the two solid curves with the CW mode under drive, the increase of the mode coupling $J$ leads to the distribution of the entanglement from the CW mode to the CCW mode, because they are coupled via the beam-splitter (state-swap) interaction, as witnessed by the increasing $E_{a_{\circlearrowleft}m}^{(\circlearrowright)}$ (the declining $E_{a_{\circlearrowright}m}^{(\circlearrowright)}$ ) with the coupling $J$ in Fig.~\ref{fig4}(a).  By contrast, for the two dashed curves with the CCW mode under drive, the CCW mode-magnon coupling $g_\circlearrowleft^{ }=0.1 g_\circlearrowright^{ }$ is too weak (under the same power) to efficiently drive the magnon mode to obtain a sufficiently strong coupling $G_m$ for creating entanglement. Although the mode coupling also leads to the driving of the CW mode, this driving is typically weak and also inefficient to achieve a strong $G_m$ for generating considerable entanglement. Consequently, the entanglements $E_{a_{\circlearrowleft}m}^{(\circlearrowleft)}$ and $E_{a_{\circlearrowright}m}^{(\circlearrowleft)}$ are either zero or negligibly small.   Similar results are observed for the microwave-phonon entanglement as shown in  Fig.~\ref{fig4}(b). We therefore conclude that the nonreciprocity of both the microwave-magnon and -phonon entanglements is robust against the backscattering induced mode coupling, even when the chiral coupling condition $g_\circlearrowright^{ }\gg g_\circlearrowleft^{ }=0$ is slightly broken. 

%With respect to the increasing of the cavity modes coupling strength $J$, the established cavity-magnon (phonon) entanglement $E_{a_{\circlearrowright}m}^{(j)}$ and $E_{a_{\circlearrowleft}m}^{(j)}$ [$E_{a_{\circlearrowright}b}^{(j)}$ and $E_{a_{\circlearrowleft}b}^{(j)}$] are provided, where the upper subscript $j=\circlearrowright,\circlearrowleft$ labels the circulating microwave cavity mode which gets direct driving. Specifically, our nonreciprocal entanglement protocol based on the chiral cavity-magnon coupling is immune to the backscattering coupling $J$, even in case that the chiral coupling condition $g_\circlearrowright^{ }\gg g_\circlearrowleft^{ }=0$ experiences slight breaking. 

Figure~\ref{fig4}(c) studies the nonreciprocal microwave-magnon entanglement versus the coupling ratio $\chi$ for a fixed $g_\circlearrowright^{ }$ with the presence of the mode coupling $J\ne0$. Clearly, the CW mode-magnon entanglement $E_{a_{\circlearrowright}m}^{(\circlearrowright)}$ (thick solid curve) is much stronger than the other three entanglements even for a sufficiently large coupling $g_\circlearrowleft^{ }{=}\,0.2 g_\circlearrowright^{ }{=}\,2\pi\times0.8$ MHz, indicating that the nonreciprocity of the entanglement is robust against the residual coupling $g_\circlearrowleft^{ }\ne0$ to the CCW mode. Such robustness is also found in the nonreciprocal microwave-phonon entanglement as manifested in Fig.~\ref{fig4}(d).

\begin{figure}[h] 
	\centering
	\includegraphics[width=1\linewidth]{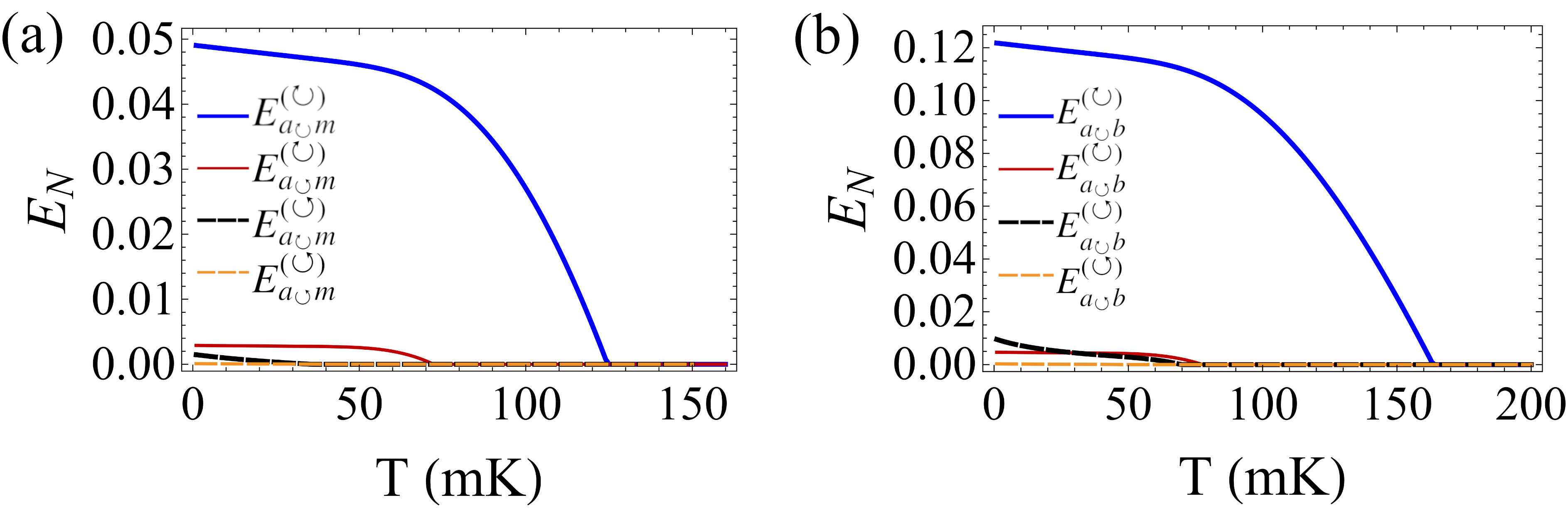}
	\caption{Stationary  entanglement $E_{a_{\circlearrowright}\nu}^{(\circlearrowright)}$ (thick solid), $E_{a_{\circlearrowleft}\nu}^{(\circlearrowright)}$ (thin solid), $E_{a_{\circlearrowright}\nu}^{(\circlearrowleft)}$ (thick dashed) and $E_{a_{\circlearrowleft}\nu}^{(\circlearrowleft)}$ (thin dashed) versus bath temperature $T$. Here, $\nu=m$ in (a) for microwave-magnon entanglement; $\nu=b$ in (b) for microwave-phonon entanglement. We take $\chi=0.1$ and the other parameters of (a) and (b) are the same as in Figs.~\ref{fig4}(c) and \ref{fig4}(d), respectively. } 
	\label{fig5}
\end{figure}	

In Fig.~\ref{fig5}, we study the effect of thermal noises on the nonreciprocal entanglements. In general,  the lower the bath temperature is, the stronger the entanglements are. Nevertheless, too low temperature also leads to the emergence of the other undesired entanglements. One thus can raise a bit the temperature, e.g., around 80 mK, to completely kill the unwanted entanglements, at the price of slightly reducing the desired entanglement.  The nonreciprocal cavity-magnon and -phonon entanglements can be achieved for the temperature up to 125~mK and 163~mK, respectively, as shown in Figs.~\ref{fig5}(a) and \ref{fig5}(b).

\begin{figure}[h] 
	\centering
	\includegraphics[width=1\linewidth]{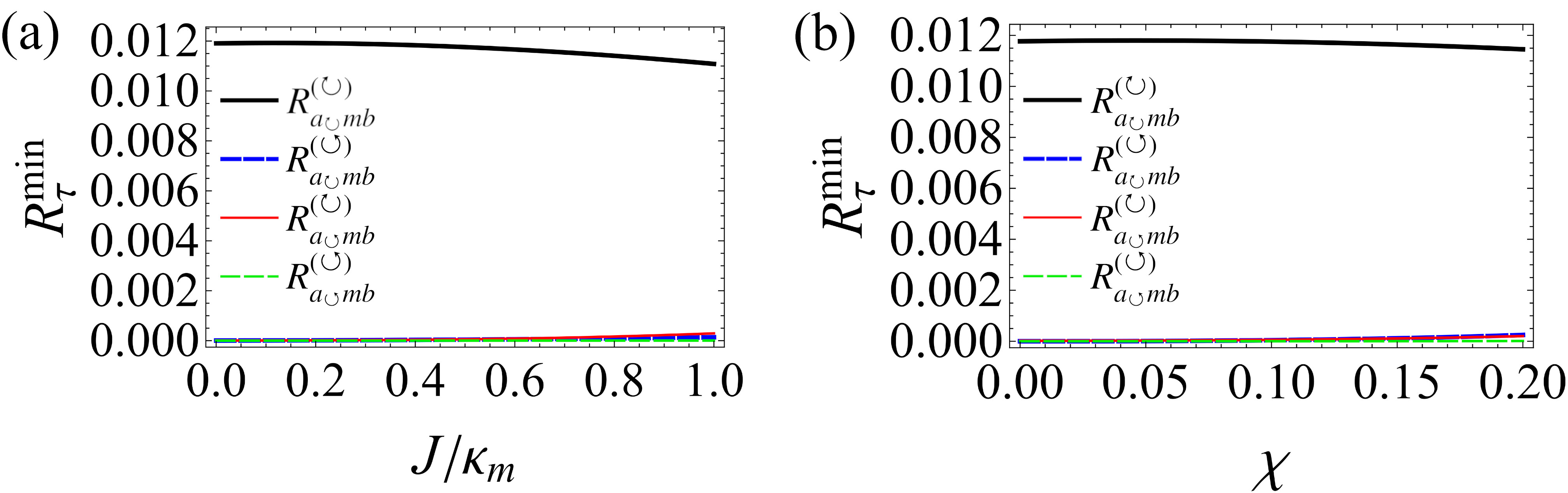}
	\caption{Tripartite entanglement in terms of the minimum residual contangle $R_{a_{\circlearrowright}mb}^{(\circlearrowright)}$ (thick solid), $R_{a_{\circlearrowleft}mb}^{(\circlearrowright)}$ (thin solid), $R_{a_{\circlearrowright}mb}^{(\circlearrowleft)}$ (thick dashed) and $R_{a_{\circlearrowleft}mb}^{(\circlearrowleft)}$ (thin dashed) versus coupling strength $J$ under $\chi =0.1$ in (a); versus $\chi$ under $J=0.5\kappa_m$ in (b). The superscript ($\circlearrowright$) or ($\circlearrowleft$) denotes the specific circulating mode under drive. The other parameters of (a) and (b) are the same as in Figs.~\ref{fig4}(a) and \ref{fig4}(c), respectively. } 
	\label{fig6}
\end{figure}	

In Fig. \ref{fig6}, we study the nonreciprocal tripartite entanglements versus the mode coupling $J$ (Fig. \ref{fig6}(a)) and the coupling ratio $\chi$ (Fig. \ref{fig6}(b)). 
The four curves represent the tripartite entanglement of the magnon mode, the mechanical mode, and the CW or the CCW mode for the two situations when the CW and CCW modes are respectively driven.   The nonreciprocity of the tripartite entanglement is clearly shown to be robust against the mode coupling $J$ and the residual coupling $g_\circlearrowleft^{ }$ to the CCW mode, as witnessed by the vanishing residual contangle $R_\tau^{\mathrm{min}}\approx 0$ of the three unwanted tripartite entanglements.

\section{Potential applications}\label{appl}

\begin{figure}[h] 
	\centering
	\includegraphics[width=0.95\linewidth]{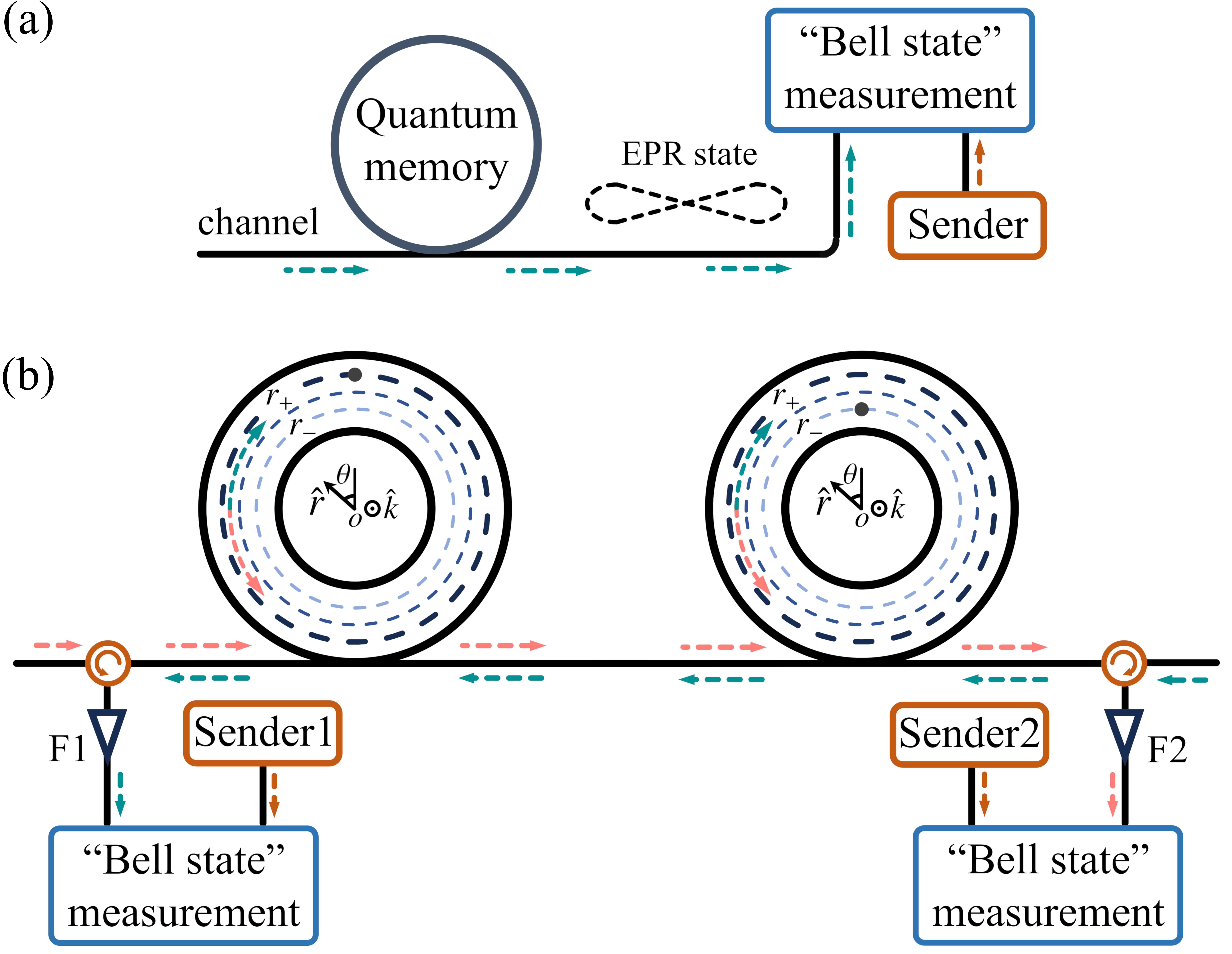}
	\caption{{(a) Schematic diagram of the conventional CV quantum teleportation. By exploiting the prepared Einstein-Podolsky-Rosen (EPR) state, the ``Bell-state" measurement (realized via the homodyne detection) teleports the quantum state from the sender to the quantum memory~\cite{cvtele}.  (b) The channel multiplexing quantum teleportation protocol based on the nonreciprocal entanglement proposed in this work. See text for description. }} 
	\label{fig7}
\end{figure}

The proposed nonreciprocal entanglements find many potential applications, one of which is  to realize the channel multiplexing quantum teleportation, as depicted in Fig.~\ref{fig7}. In the conventional continuous-variables (CV) quantum teleportation scheme (Fig.~\ref{fig7}(a))~\cite{cvtele}, the teleportation of a quantum state from the sender onto the quantum memory exploits a single channel. By contrast, the nonreciprocal entanglement allows for a channel multiplexing protocol (Fig.~\ref{fig7}(b)), where the chiral cavity-magnon coupling enables the microwave photons injecting from the left (right) side to couple with the magnomechanical system in the right (left) cavity and entangle with the magnons/phonons contained therein, but decouple from the magnomechanical system in the left (right) cavity.  This allows for the teleportation of the quantum state from the senders at both sides onto their respective quantum memories still using a single channel, thereby increasing the usage efficiency of the channel compared to the conventional teleportation scheme.

We now discuss the performance of such a channel multiplexing protocol based on, e.g., the generated nonreciprocal microwave-magnon entanglement, and calculate the corresponding teleportation state fidelity. In the CV quantum teleportation, the fidelity of teleporting an input single-mode Gaussian state by exploiting a two-mode Gaussian entangled state is given by \cite{Fiurasek2002}
\begin{equation} \label{fidelity}
	\mathcal{F}=\frac{1}{\sqrt{\mathrm{det}V}},
\end{equation}
where $V=2V_{in}+\sigma_z V_e \sigma_z^T + \sigma_z V_{ef} + V_{ef}^T\sigma_z^T + V_f$, with $\sigma_z=\mathrm{diag}[1,-1]$ being the Pauli-$z$ matrix, $V_{in}$ being the CM of the input state, and $V_e$, $V_f$, $V_{ef}$ and $V_{fe}$ being $2\times 2$ blocks of the CM of the two-mode entangled state. To be specific, we consider a microwave coherent state as the input state with its CM $V_{in}=\mathrm{diag}[1/2,1/2]$, and the generated microwave-magnon entangled state as the two-mode entangled state. It should be noted that the above microwave-magnon entanglement is between the microwave {\it cavity mode} and the magnon mode, while the subsequent ``Bell-state" measurement is applied onto the cavity output field and the input coherent state. To estimate the fidelity using Eq.~\eqref{fidelity}, one has to properly define the {\it cavity output mode} to have a physically valid CM.  The cavity output mode can be defined with a filter function with a properly chosen central frequency $\Omega$ and bandwidth $1/\tau$ (F1 and F2 in Fig. \ref{fig7}(b) denote two filters), given by~\cite{Genes2008}
\begin{equation}
	\tilde{g}(\omega) = \sqrt{\frac{\tau}{2\pi}}e^{i(\omega-\Omega)\tau/2}\frac{\mathrm{sin}[(\omega-\Omega)\tau/2]}{(\omega-\Omega)\tau/2}.
\end{equation}
This then allows us to calculate the corresponding CM of the cavity output mode and the magnon mode, which are entangled.

Under the parameters of Fig. \ref{fig2}(d) and taking $\gamma_b/2\pi=10^2$ Hz, we obtain the entanglement between the cavity output mode and the magnon mode $E_{\mathrm{am}}=0.23$, which leads to the teleportation state fidelity $\mathcal{F}=0.55$, by choosing the central frequency $\Omega=-\omega_b$ (with respect to $\omega_0$, cf. Fig.~\ref{fig1}(b)) and an optimal bandwidth $\tau^{-1}\approx0.1\omega_b$~\cite{Genes2008} of the filter.  The state fidelity $\mathcal{F}=0.55$ is above the minimum threshold of 0.5 for quantum teleportation with a coherent input state~\cite{Braunstein2000}. We remark that, although the state fidelity can be high, e.g., up to 0.83, for CV quantum teleportation in purely optical systems~\cite{Yukawa2008,Pirandola2015,Hu2023}, the fidelity in the CV quantum teleportation between objects of a different nature (e.g., light and matter) is typically much lower: the fidelity in the range of $0.58{-}0.60$ was demonstrated in the teleportation from light to an atomic ensemble~\cite{Polzik2006}. Our fidelity $\mathcal{F}=0.55$ in the microwave-to-magnon quantum teleportation is comparable to the above values.

%Clearly, by properly filtering the output field, one realizes an effective entanglement distillation because the selected output mode is more entangled than the cavity mode with the magnon mode.

Besides the applications in the channel multiplexing quantum teleportation, the work may be exploited for designing controllable quantum devices. In the chiral CMM system as illustrated in Fig.~\ref{fig1}(a), the transmission direction of the circulating microwave cavity mode entangled with the magnon (phonon) mode can be conveniently switched on demand by reversing the bias magnetic field to change the precession direction of the magnetization~\cite{Xie2023}. Such fast switch entanglement may find applications in relevant quantum protocols.

\section{Conclusion}\label{conc}
We present a scheme for achieving nonreciprocal microwave -magnon and -phonon bipartite entanglements and photon-magnon-phonon tripartite entanglement in a CMM system based on the chiral cavity-magnon coupling. This is realized in a torus-shaped microwave cavity by placing the YIG sphere on a special line in the cavity to break the time-reversal symmetry of the two degenerate counter-propagating cavity modes. We analyze the effects of some major experimental imperfections, such as the coupling between the two degenerate cavity modes and the simultaneous couplings of the magnon mode to the two degenerate modes, and find that the nonreciprocity of the entanglements exhibits strong robustness towards these imperfections. Furthermore, we specifically provide a channel multiplexing quantum teleportation scheme based on the nonreciprocal entanglement. The work may find promising applications of the cavity magnomechanical systems in noise-tolerant quantum processing, channel multiplexing quantum teleportation, and chiral magnonic quantum networks.

\appendix
	\section*{Appendix A: Methods to verify magnon frequency combs}
	A strong driving field can generate magnon frequency combs \cite{Xiong23FR}. To verify if this is present under the parameters we achieve entanglement, in the following we solve the differential equations to study the time evolution of the average of each mode. For completeness, here we include the effect of the backscattering induced coupling between the two degenerate counter-propagating cavity modes as in Sec. \ref{imperfect}. The Hamiltonian of the system becomes
	\begin{equation} \label{wholeHamil}
		\begin{split}
			&H'/\hbar=\\
			&\sum_{j=\circlearrowright,\circlearrowleft}\omega_a a_j^\dagger a_j^{ }+\omega_m m^\dagger m+\frac{\omega_b}{2}(q^2+p^2)+g_{m} m^\dagger m q\\
			&+ \sum_{j=\circlearrowright,\circlearrowleft} g_j^{ } (a_j^\dagger m+a_j m^\dagger)+J(a_\circlearrowright^{ } a_\circlearrowleft^\dagger + a_\circlearrowright^\dagger a_\circlearrowleft^{ })+H_d/\hbar,
		\end{split}
	\end{equation}
	where $J$ is the coupling strength between the two degenerate modes. When driving  the CW mode of the cavity, the Hamiltonian \eqref{wholeHamil} gives rise to the following differential equations for the classical averages of the modes
	\begin{equation}
		\begin{split}
			\dot{\langle a_\circlearrowright^{ }\rangle}=&-i\Delta_{\circlearrowright} \langle a_\circlearrowright^{ }\rangle-\kappa_a \langle a_\circlearrowright^{ }\rangle -iJ\langle a_\circlearrowleft^{ }\rangle-i g_\circlearrowright^{ } \langle m\rangle+ E,\\
			\dot{\langle a_\circlearrowleft^{ }\rangle }=&-i\Delta_{\circlearrowleft} \langle a_\circlearrowleft^{ }\rangle-\kappa_a \langle a_\circlearrowleft^{ }\rangle -iJ\langle a_\circlearrowright^{ }\rangle -i g_\circlearrowleft^{ } \langle m\rangle,\\
			\dot{\langle m\rangle }=&-i\Delta_m \langle m\rangle-\kappa_m \langle m\rangle-i g_m \langle m\rangle \langle q\rangle\\
			&-i g_\circlearrowright^{ } \langle a_\circlearrowright^{ }\rangle -i g_\circlearrowleft^{ } \langle a_\circlearrowleft^{ }\rangle,\\
			\dot{\langle q\rangle}=&\ \omega_b \langle p\rangle,\ \ 
			\dot{\langle p\rangle}=-\omega_b \langle q\rangle -\gamma_b \langle p\rangle -g_m \langle m^\dagger\rangle \langle m\rangle.\\
		\end{split}
	\end{equation}
	By solving the above equations and studying the time evolution of each mode, we can verify if the system enters a parameter regime where magnon frequency combs are generated.
	
	\begin{figure}[h] 
		\centering
		\includegraphics[width=0.75\linewidth]{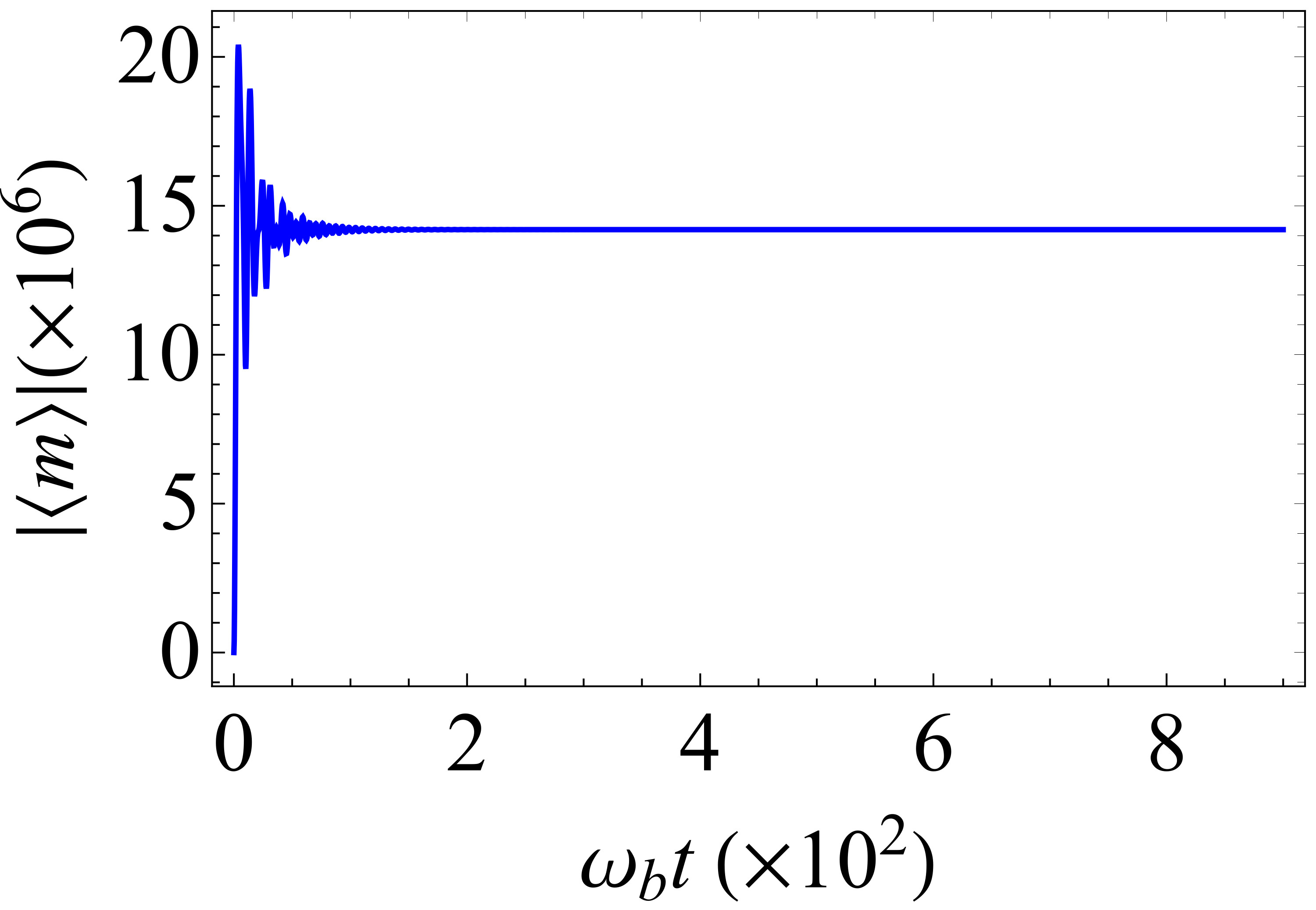}
		\caption{The time evolution of the magnon amplitude $|\langle m(t)\rangle|$. Here we take $J=0.5\kappa_m$ and the other parameters are the same as used in Fig. \ref{fig4}(a).} 
		\label{figs1}
	\end{figure}

	Under the parameters of Fig. \ref{fig4}(a) and taking $J=0.5\kappa_m$, combining the corresponding initial conditions, we solve the time evolution of each mode, and Fig. \ref{figs1} shows the result for the magnon mode amplitude $|\langle m(t)\rangle|$. It shows that the magnon amplitude evolves towards a stable value (similar to Fig. 2(a) of Ref. \cite{Xiong23FR}), which implies that the drive power we used is below the threshold for producing magnon frequency combs. So, the presence of the frequency combs is excluded.

\section*{Appendix B: Linearized QLEs considering experimental imperfections}

The Hamiltonian including the coupling between the two degenerate counter-propagating cavity modes is shown in Eq. \eqref{wholeHamil}. For the case where the microwave drive is applied to the CW circulating cavity mode, the linearized QLEs describing the quadrature fluctuations of the system can be written in the matrix form of 
\begin{equation}
	\dot{u}(t)=A'u(t)+n(t),
\end{equation}
where the vectors $u(t)$ and $n(t)$ are the same as in Eq.~\eqref{matrixform}, but the drift matrix $A'$ is updated as follows:
\begin{small}
	\begin{equation} \label{renewdrift}
		\begin{split}
			&A'=\\
			&\begin{pmatrix}
				-\kappa_a & \Delta_\circlearrowright & 0 & J & 0 & g_\circlearrowright^{ } & 0 & 0\\
				-\Delta_\circlearrowright & -\kappa_a & -J & 0 & -g_\circlearrowright^{ } & 0 & 0 & 0\\
				0 & J & -\kappa_a & \Delta_\circlearrowleft & 0 & g_\circlearrowleft^{ } & 0 & 0\\
				-J & 0 & -\Delta_\circlearrowleft & -\kappa_a & -g_\circlearrowleft^{ } & 0 & 0 & 0\\
				0 & g_\circlearrowright^{ } & 0 & g_\circlearrowleft^{ } & -\kappa_m & \tilde{\Delta}_m & \mathrm{Im}G_m   & 0\\
				-g_\circlearrowright^{ } & 0 & -g_\circlearrowleft^{ } & 0 & -\tilde{\Delta}_m & -\kappa_m & -\mathrm{Re}G_m & 0\\
				0 & 0 & 0 & 0 & 0 & 0 & 0 & \omega_b\\
				0 & 0 & 0 & 0 & -\mathrm{Re}G_m & -\mathrm{Im}G_m & -\omega_b & -\gamma_b
			\end{pmatrix},
		\end{split}
	\end{equation}
\end{small}
where the effective magnomechanical coupling $G_m{=}\sqrt{2}g_m\langle m\rangle$ is redefined with the following new steady-state average:
\begin{equation}
	\begin{split}
		&\langle m\rangle=\\
		&\frac{E[(g_\circlearrowright^{ }\Delta_{a}-g_\circlearrowleft^{ }J)-i g_\circlearrowright^{ }\kappa_a]}{(\kappa_m J^2+\kappa_a \epsilon_1-\Delta_{a}\epsilon_2)+i(\tilde{\Delta}_m J^2-2J g_\circlearrowright^{ } g_\circlearrowleft^{ }+\Delta_{a}\epsilon_1+\kappa_a\epsilon_2)},
	\end{split}
\end{equation}
with $\epsilon_1=g_\circlearrowright^2+g_\circlearrowleft^2+\kappa_a\kappa_m-\Delta_{a}\tilde{\Delta}_m$, $\epsilon_2=\kappa_a\tilde{\Delta}_m+\kappa_m\Delta_a$, $\tilde{\Delta}_m=\Delta_m+g_m\langle q\rangle$, and $\langle q\rangle=-g_m|\langle m\rangle|^2/\omega_b$.

For the case where the CCW circulating cavity mode is driven, the drift matrix takes the same form as in Eq.~\eqref{renewdrift}, while the steady-state average $\langle m\rangle$ takes a slightly different form of
\begin{equation}
	\begin{split}
		&\langle m\rangle=\\
		&\frac{E[(g_\circlearrowleft^{ }\Delta_{a}-g_\circlearrowright^{ }J)-i g_\circlearrowleft^{ }\kappa_a]}{(\kappa_m J^2+\kappa_a \epsilon_1-\Delta_{a}\epsilon_2)+i(\tilde{\Delta}_m J^2-2J g_\circlearrowright^{ } g_\circlearrowleft^{ }+\Delta_{a}\epsilon_1+\kappa_a\epsilon_2)},
	\end{split}
\end{equation}
where $\epsilon_1$, $\epsilon_2$, $\tilde{\Delta}_m$ and $\langle q\rangle$ are defined in the same way as above.

%\end{appendices}
% Use the style of numbering in square brackets.
% If nothing is used, default style will be taken.
%\begin{enumerate}[a)]
%\item 
%\item 
%\item 
%\end{enumerate}  

%\begin{itemize}
%\item 
%\item 
%\item 
%\end{itemize}  

%\begin{description}
%\item[]
%\item[] 
%\item[] 
%\end{description}  

% Uncomment and use as the case may be
%\begin{theorem} 
%\end{theorem}

% Uncomment and use as the case may be
%\begin{lemma} 
%\end{lemma}



\section*{Acknowledgements}
We thank H. Jing for valuable feedback on the manuscript, and Z.-Q. Wang and R.-C. Shen for helpful discussions. This work was supported by National Key Research and Development Program of China (Grant No. 2022YFA1405200, 2024YFA1408900), National Natural Science Foundation of China (Grant No. 12474365, 92265202), and Zhejiang Provincial Natural Science Foundation of China (Grant No. LR25A050001).

\section*{Declaration of competing interest}
The authors declare that they have no conflict of interest.

%\bibliographystyle{cas-model2-names}

%\bibliography{cas-refs}



\end{document}